\documentclass[11pt]{article}
\usepackage{amsmath,amsfonts,latexsym,graphicx,amssymb}
\usepackage{amsfonts}
\usepackage{bm}
\newcommand{\ot}{{\,\otimes\,}}
\newcommand{{\Cd}}{{\mathbb{C}^3}}

\def\oper{{\mathchoice{\rm 1\mskip-4mu l}{\rm 1\mskip-4mu l}{\rm 1\mskip-4.5mu l}{\rm 1\mskip-5mu l}}}
\def\<{\langle}
\def\>{\rangle}

\begin{document}

\date{}
\title{\textbf{On the symmetry of the seminal Horodecki state}}

\author{Dariusz  Chru\'sci\'nski and
Andrzej Kossakowski \\
Institute of Physics, Nicolaus Copernicus University,\\
Grudzi\c{a}dzka 5/7, 87--100 Toru\'n, Poland}

\maketitle

\begin{abstract}
It is shown that the seminal Horodecki 2-qutrit state belongs to the
class of states displaying symmetry governed by a commutative
subgroup of the unitary group $U(3)$. Taking a conjugate subgroup
one obtains another classes of symmetric states and one finds
equivalent representations of the Horodecki state.
\end{abstract}

\section{Introduction}

%\noindent {\it Introduction}.

In a seminal paper \cite{Pawel} Pawe{\l} Horodecki provided an
example of a density operator living in $\Cd \ot \Cd$ which
represents entangled state positive under partial transposition
(PPT)
\begin{equation}\label{RHO}
    \rho_a\, =\, N_a\left(\begin{array}{ccc|ccc|ccc}
   a&\cdot&\cdot&\cdot&a&\cdot&\cdot&\cdot&a\\
   \cdot&a&\cdot&\cdot&\cdot&\cdot&\cdot&\cdot&\cdot\\
   \cdot&\cdot&a&\cdot&\cdot&\cdot&\cdot&\cdot&\cdot\\
   \hline
   \cdot&\cdot&\cdot&a&\cdot&\cdot&\cdot&\cdot&\cdot\\
   a&\cdot&\cdot&\cdot&a&\cdot&\cdot&\cdot&a\\
   \cdot&\cdot&\cdot&\cdot&\cdot&a&\cdot&\cdot&\cdot\\
   \hline
   \cdot&\cdot&\cdot&\cdot&\cdot&\cdot&b&\cdot&c\\
   \cdot&\cdot&\cdot&\cdot&\cdot&\cdot&\cdot&a&\cdot\\
   a&\cdot&\cdot&\cdot&a&\cdot&c&\cdot&b
   \end{array}\right) \ ,
\end{equation}
with
\begin{equation}\label{}
    N_a = \frac{1}{8a+1}\ ,\ \ \ \ b = \frac{1+a}{2}\ , \ \ \ \ c = \frac{\sqrt{1-a^2}}{2}\ ,
\end{equation}
and $a \in [0,1]$. The above matrix representation corresponds to
the standard computational basis $|ij\> = |i\> \ot |j\>$ in $\Cd\ot
\Cd$ and to make the picture more transparent we replaced all zeros
by dots. Since the partial transposition $\rho_a^\Gamma = (\oper \ot
{\rm T})\rho_a \geq 0$ the state is PPT for all $a\in [0,1]$. It is
easy to show that for $a=0$ and $a=1$ the state is separable and it
was shown \cite{Pawel} that for $a \in (0,1)$ the state is entangled
(for the recent reviews of quantum entanglement and the methods of
its detection see \cite{HHHH} and \cite{Guhne}). Actually, the
family (\ref{RHO}) provides one of the first examples of bound
entanglement. In this Letter we analyze the structure of
(\ref{RHO}). In particular we study its symmetry group.

%\vspace{.5cm}

\section{Symmetry group}

Let $G$ be a subgroup of the unitary group $U(d)$ (a group of
unitary $d \times d$ matrices).   A state $\rho$ living in
$\mathbb{C}^d \ot \mathbb{C}^d$ is $G \ot \overline{G}$--invariant
if
\begin{equation}\label{iso}
    U \ot \overline{U} \rho = \rho\,  U \ot \overline{U} \ ,
\end{equation}
where $U \in G$, and $\overline{U}$ denotes the complex conjugation
of the matrix elements with respect to the computational basis
$|i\>$. It is clear that if $\rho$ is $G \ot
\overline{G}$--invariant then its partial transposition is $G \ot
G$--invariant, that is
\begin{equation}\label{werner}
    U \ot {U} \rho = \rho\,  U \ot U \ ,
\end{equation}
where $U \in G$. Recall, that if $G = U(d)$, then $G \ot
\overline{G}$--invariant states define a class of isotropic states
\cite{ISO}, whereas $G \ot G$--invariant states define a class of
Werner states \cite{Werner1}. Recently \cite{PPT-nasza} we found a
class of $G \ot \overline{G}$--invariant states, where $G$ defines a
maximal abelian subgroup of $U(d)$ defined as follows:
\begin{equation}\label{U-x}
U_\mathbf{x} = \exp\left(i \sum_{k=1}^{d} x_k |k\>\<k| \right)\ ,
\end{equation}
and $\mathbf{x}=(x_1,\ldots,x_{d}) \in \mathbb{R}^d$. It was shown
\cite{PPT-nasza} that states invariant under the maximal abelian
subgroup have the following structure
\begin{equation}\label{}
    \rho = \sum_{i,j=1}^d a_{ij}\, |i\>\<j| \ot |i\>\<j| + \sum_{i\neq
    j=1}^d d_{ij}\, |i\>\<i| \ot |j\>\<j|\ ,
\end{equation}
where the matrix $||a_{ij}|| \geq 0$, and the numbers $d_{ij} \geq
0$. The normalization condition gives
\[  \sum_{i=1}^d a_{ii} + \sum_{i\neq  j=1}^d d_{ij}      = 1 \ . \]
The corresponding matrix representation for $d=3$ reads as follows
\begin{equation}\label{RHO-abelian}
    \rho\, =\, \left(\begin{array}{ccc|ccc|ccc}
   a_{11}&\cdot&\cdot&\cdot&a_{12}&\cdot&\cdot&\cdot&a_{13}\\
   \cdot&d_{12}&\cdot&\cdot&\cdot&\cdot&\cdot&\cdot&\cdot\\
   \cdot&\cdot&d_{13}&\cdot&\cdot&\cdot&\cdot&\cdot&\cdot\\
   \hline
   \cdot&\cdot&\cdot&d_{21}&\cdot&\cdot&\cdot&\cdot&\cdot\\
   a_{21}&\cdot&\cdot&\cdot&a_{22}&\cdot&\cdot&\cdot&a_{23}\\
   \cdot&\cdot&\cdot&\cdot&\cdot&d_{23}&\cdot&\cdot&\cdot\\
   \hline
   \cdot&\cdot&\cdot&\cdot&\cdot&\cdot&d_{31}&\cdot&\cdot\\
   \cdot&\cdot&\cdot&\cdot&\cdot&\cdot&\cdot&d_{32}&\cdot\\
   a_{31}&\cdot&\cdot&\cdot&a_{32}&\cdot&\cdot&\cdot&a_{33}
   \end{array}\right) \ .
\end{equation}
Let us observe that (\ref{RHO-abelian}) is PPT if and only if
\begin{equation}\label{}
    d_{ij} d_{ji} \geq |a_{ij}|^2\ , \ \ \ \ i\neq j\ .
\end{equation}
Surprisingly many well know states considered in the literature
belong to this class (see \cite{PPT-nasza} for examples). Note,
however, that  Horodecki state (\ref{RHO}) does not belong to
(\ref{RHO-abelian}) unless $a=1$. Consider now a subgroup $G_0$ of
the $G$ defined by (\ref{U-x}) with $x_1=x_3$. One finds the
following structure of invariant states
\begin{equation}  \label{RHO-13}
  \rho = \left( \begin{array}{ccc|ccc|ccc}
\rho_{11} & \cdot & \rho_{13}  & \cdot & \rho_{15} & \cdot &  \rho_{17} & \cdot &  \rho_{19} \\
\cdot & \rho_{22} & \cdot & \cdot & \cdot & \cdot & \cdot &
\rho_{28} &
\cdot \\
\rho_{31} & \cdot & \rho_{33}  & \cdot & \rho_{35} & \cdot &
\rho_{37} & \cdot & \rho_{39} \\ \hline \cdot & \cdot & \cdot &
\rho_{44} & \cdot & \rho_{46} & \cdot & \cdot &
\cdot \\
\rho_{51} & \cdot & \rho_{53} &  \cdot & \rho_{55} & \cdot &
\rho_{57} & \cdot &
\rho_{59} \\
\cdot & \cdot & \cdot & \rho_{64} & \cdot &  \rho_{66}  & \cdot & \cdot & \cdot \\
\hline
\rho_{71} & \cdot & \rho_{73}  & \cdot & \rho_{75} & \cdot &  \rho_{77} & \cdot &  \rho_{79} \\
\cdot & \rho_{82} & \cdot & \cdot & \cdot & \cdot & \cdot &
\rho_{88} &
\cdot \\
\rho_{91} & \cdot & \rho_{93}  & \cdot & \rho_{95} & \cdot &
\rho_{97} & \cdot & \rho_{99}
\end{array} \right) \ ,
\end{equation}
and it evidently contains Horodecki state (\ref{RHO}).
Interestingly, invariant states (\ref{RHO-13}) have almost perfect
chessboard structure \cite{Bruss} (see also the recent paper
\cite{Djokovic}.  Note, however, that only a subclass of states
considered in \cite{Bruss,Djokovic} are $G_0 \ot G_0$--invariant.
The characteristic feature of (\ref{RHO-13}) is that $\rho$ has a
direct sum structure $\rho = \rho_1 \oplus \rho_2 \oplus \rho_3$\,
where the corresponding operators $\rho_k$ are supported on
$\mathcal{H}_k$
\begin{eqnarray} \label{3H}
  \mathcal{H}_1 &=& \rm{span}_{\,\mathbb{C}} \{\, |11\>,\ |13\>,\ |22\>,\ |31\>,\ |33\>\, \} \ , \nonumber\\
  \mathcal{H}_2 &=&  \rm{span}_{\,\mathbb{C}} \{\, |12\>,\ |32\>\, \} \ , \\
  \mathcal{H}_3 &=& \rm{span}_{\,\mathbb{C}} \{\, |21\>,\ |23\>\, \} \nonumber  \
  ,
\end{eqnarray}
giving rise to  the direct sum decomposition
%\begin{equation}\label{}
 $   \mathbb{C}^3 \ot \mathbb{C}^3 = \mathcal{H}_1 \oplus
    \mathcal{H}_2 \oplus \mathcal{H}_3$.
%\end{equation}
Similarly, the partial transposition
\begin{equation}\label{Gamma}
 \rho^\Gamma = \left( \begin{array}{ccc|ccc|ccc}
\rho_{11} & \cdot & \rho_{31}  & \cdot & \cdot & \cdot &  \rho_{17} & \cdot &  \rho_{37} \\
\cdot & \rho_{22} & \cdot & \rho_{15} & \cdot & \rho_{35} & \cdot &
\rho_{28} &
\cdot \\
\rho_{13} & \cdot & \rho_{33}  & \cdot & \cdot & \cdot & \rho_{19} &
\cdot & \rho_{39} \\ \hline \cdot & \rho_{51} & \cdot & \rho_{44} &
\cdot & \rho_{64} & \cdot & \rho_{57} &
\cdot \\
\cdot & \cdot & \cdot &  \cdot & \rho_{55} & \cdot & \cdot & \cdot &
\cdot \\
\cdot & \rho_{53} & \cdot & \rho_{46} & \cdot &  \rho_{66}  & \cdot & \rho_{59} & \cdot \\
\hline
\rho_{71} & \cdot & \rho_{91}  & \cdot & \cdot & \cdot &  \rho_{77} & \cdot &  \rho_{97} \\
\cdot & \rho_{82} & \cdot & \rho_{75} & \cdot & \rho_{95} & \cdot &
\rho_{88} &
\cdot \\
\rho_{73} & \cdot & \rho_{93}  & \cdot & \cdot & \cdot & \rho_{79} &
\cdot & \rho_{99}
\end{array} \right)
\end{equation}
has a direct sum structure $\rho^\Gamma = \widetilde{\rho}_1 \oplus
\widetilde{\rho}_2 \oplus \widetilde{\rho}_3$\, where the
corresponding operators $\widetilde{\rho}_k$ are supported on
$\widetilde{\mathcal{H}}_k$
\begin{eqnarray} \label{3Ha}
  \widetilde{\mathcal{H}}_1 &=& \rm{span}_{\,\mathbb{C}} \{\, |11\>,\ |13\>,\ |31\>,\ |33\>\, \} \ , \nonumber\\
  \widetilde{\mathcal{H}}_2 &=&  \rm{span}_{\,\mathbb{C}} \{\, |12\>,\ |21\>,\ |23\>,\ |32\>\, \} \ , \\
  \widetilde{\mathcal{H}}_3 &=& \rm{span}_{\,\mathbb{C}} \{\, |22\>\, \} \nonumber  \
  ,
\end{eqnarray}
together with $   \mathbb{C}^3 \ot \mathbb{C}^3 =
\widetilde{\mathcal{H}}_1 \oplus \widetilde{\mathcal{H}}_2 \oplus
\widetilde{\mathcal{H}}_3$. Interestingly one has
\begin{equation}\label{HHHH}
    \widetilde{\mathcal{H}}_1 \oplus \widetilde{\mathcal{H}}_3 =
    \mathcal{H}_1\ , \ \ \ \mathcal{H}_2 \oplus \mathcal{H}_3 =
    \widetilde{\mathcal{H}}_2\ .
\end{equation}
Hence to check for PPT one needs to check positivity of two $4\times
4$ leading submatrices of (\ref{Gamma}). Note, that decompositions
(\ref{3H}) and (\ref{3Ha}) remind the characteristic circulant
decompositions \cite{CIRCULANT}. There is however important
difference: (\ref{3H}) and (\ref{3Ha}) are governed by the symmetry
group $G_0$ whereas the circulant decompositions are not directly
related to any symmetry. For other types of decompositions which
simplify PPT conditions see also \cite{Brazylia}.

\section{Another representations of the Horodecki state}

Consider now another commutative subgroup $G_0'$ defined by
$x_1=x_2$. It is clear that
\begin{equation}\label{}
    G_0' = S' G_0 S'^\dagger\ ,
\end{equation}
where $S'$ represents permutation $(1,2,3) \rightarrow (1,3,2)\,$,
that is
\begin{equation}\label{}
    S' = \left( \begin{array}{ccc} 1 & \cdot & \cdot \\ \cdot & \cdot & 1 \\
    \cdot & 1 & \cdot \end{array} \right) \ .
\end{equation}
Hence a class of $G_0' \ot \overline{G}_0'$--invariant states is
defined by
\begin{equation}\label{}
\rho' = S' \ot S'\, \rho\, S'^\dagger \ot S'^\dagger\ ,
\end{equation}
where $\rho$ is  $G_0 \ot \overline{G}_0$--invariant. The
corresponding matrix representation  of $\rho'$ has the following
form
\begin{equation}\label{1=2}
 \rho' = \left( \begin{array}{ccc|ccc|ccc}
\rho_{11} & \rho_{12} & \cdot &  \rho_{14} & \rho_{15} & \cdot &  \cdot & \cdot &  \rho_{19} \\
\rho_{21} & \rho_{22} & \cdot & \rho_{24} & \rho_{25} & \cdot &
\cdot & \cdot &
\rho_{29} \\
\cdot & \cdot &   \rho_{33} & \cdot & \cdot & \rho_{36} & \cdot & \cdot & \cdot \\
\hline \rho_{41} & \rho_{42} & \cdot & \rho_{44} & \rho_{45} & \cdot
& \cdot & \cdot & \rho_{49}
\\
\rho_{51} & \rho_{52} & \cdot & \rho_{54} & \rho_{55} & \cdot &
\cdot & \cdot &
\rho_{59} \\
\cdot & \cdot & \rho_{63} & \cdot & \cdot &  \rho_{66}  & \cdot & \cdot & \cdot \\
\hline \cdot & \cdot
& \cdot &  \cdot & \cdot & \cdot &  \rho_{77} &  \rho_{78} &  \cdot \\
\cdot & \cdot
& \cdot &  \cdot & \cdot & \cdot &  \rho_{87} &  \rho_{88} &  \cdot \\
\rho_{91} & \rho_{92} &  \cdot & \rho_{94} &  \rho_{95} & \cdot &
\cdot & \cdot & \rho_{99}
\end{array} \right)\  .
\end{equation}
In particular one obtains the following representation of the
Horodecki state invariant under $G_0'$
\begin{equation}\label{}
{\rho_a}' = S' \ot S'\, \rho_a\, S'^\dagger \ot S'^\dagger\ ,
\end{equation}
or in the matrix form
\begin{equation}\label{RHO-12}
    {\rho_a}'\, =\, N_a\left(\begin{array}{ccc|ccc|ccc}
   b&c&\cdot&\cdot&a&\cdot&\cdot&\cdot&a\\
   c&b&\cdot&\cdot&\cdot&\cdot&\cdot&\cdot\\
   \cdot&\cdot&a&\cdot&\cdot&\cdot&\cdot&\cdot&\cdot\\
   \hline
   \cdot&\cdot&\cdot&a&\cdot&\cdot&\cdot&\cdot&\cdot\\
   a&\cdot&\cdot&\cdot&a&\cdot&\cdot&\cdot&a\\
   \cdot&\cdot&\cdot&\cdot&\cdot&a&\cdot&\cdot&\cdot\\
   \hline
   \cdot&\cdot&\cdot&\cdot&\cdot&\cdot&a&\cdot&\cdot\\
   \cdot&\cdot&\cdot&\cdot&\cdot&\cdot&\cdot&a&\cdot\\
   a&\cdot&\cdot&\cdot&a&\cdot&\cdot&\cdot&a
   \end{array}\right) \ .
\end{equation}
The characteristic feature of (\ref{1=2}) is that $\rho'$ has a
direct sum structure $\rho' = \rho_1' \oplus \rho_2' \oplus
\rho_3'$\, where the corresponding operators $\rho_k$ are supported
on $\mathcal{H}_k'$
\begin{eqnarray}
  \mathcal{H}_1' &=& (S'\ot S')\mathcal{H}_1 = \rm{span}_{\,\mathbb{C}} \{\, |11\>,\ |12\>,\ |21\>,\ |22\>,\ |33\>\, \} \ , \nonumber\\
  \mathcal{H}_2' &=&  (S'\ot S')\mathcal{H}_2 =\rm{span}_{\,\mathbb{C}} \{\, |13\>,\ |23\>\, \} \ , \\
  \mathcal{H}_3' &=& (S'\ot S')\mathcal{H}_3 =\rm{span}_{\,\mathbb{C}} \{\, |31\>,\ |32\>\, \} \nonumber
  \ .
\end{eqnarray}
One easily finds for the partial transposition
\begin{equation}\label{}
  {\rho'}^\Gamma = \left( \begin{array}{ccc|ccc|ccc}
\rho_{11} & \rho_{21} & \cdot &  \rho_{14} & \rho_{24} & \cdot &  \cdot & \cdot &  \cdot \\
\rho_{12} & \rho_{22} & \cdot & \rho_{15} & \rho_{25} & \cdot &
\cdot & \cdot &
\cdot \\
\cdot & \cdot &   \rho_{33} & \cdot & \cdot & \rho_{36} & \rho_{19} & \rho_{29} & \cdot \\
\hline \rho_{41} & \rho_{51} & \cdot & \rho_{44} & \rho_{54} & \cdot
& \cdot & \cdot & \cdot
\\
\rho_{42} & \rho_{52} & \cdot & \rho_{45} & \rho_{55} & \cdot &
\cdot & \cdot &
\cdot \\
\cdot & \cdot & \rho_{63} & \cdot & \cdot &  \rho_{66}  & \rho_{49} & \rho_{59} & \cdot \\
\hline \cdot & \cdot
& \rho_{91} &  \cdot & \cdot & \rho_{94} &  \rho_{77} &  \rho_{87} &  \cdot \\
\cdot & \cdot
& \rho_{92} &  \cdot & \cdot & \rho_{95} &  \rho_{78} &  \rho_{88} &  \cdot \\
\cdot & \cdot &  \cdot & \cdot &  \cdot & \cdot & \cdot & \cdot &
\rho_{99}
\end{array} \right)\ .
\end{equation}
It is evident that ${\rho'}^\Gamma$ has a direct sum structure
${\rho'}^\Gamma = \widetilde{\rho}_1' \oplus \widetilde{\rho}_2'
\oplus \widetilde{\rho}_3'$\, where the corresponding operators
$\widetilde{\rho}_k'$ are supported on $\mathcal{H}_k'$
\begin{eqnarray}
  \widetilde{\mathcal{H}}_1'  &=& (S'\ot S')\widetilde{\mathcal{H}}_1 = \rm{span}_{\,\mathbb{C}} \{\, |11\>,\ |12\>,\ |21\>,\ |22\>,\, \} \ , \nonumber\\
  \widetilde{\mathcal{H}}_2' &=&  (S'\ot S')\widetilde{\mathcal{H}}_2 = \rm{span}_{\,\mathbb{C}} \{\, |13\>,\ |23\>, \, |31\>,\ |32\>\, \} \ , \\
  \widetilde{\mathcal{H}}_3' &=& (S'\ot S')\widetilde{\mathcal{H}}_3 = \rm{span}_{\,\mathbb{C}} \{\, |33\>\, \} \nonumber
  \ .
\end{eqnarray}
Again the analog of the formulae (\ref{HHHH}) holds, that is
\begin{equation}\label{HHHH-prime}
    \widetilde{\mathcal{H}}_1' \oplus \widetilde{\mathcal{H}}_3' =
    \mathcal{H}_1'\ , \ \ \ \mathcal{H}_2' \oplus \mathcal{H}_3' =
    \widetilde{\mathcal{H}}_2'\ .
\end{equation}
Finally, let us consider another commutative subgroup $G_0''$ of $G$
defined by $x_2=x_3$. It is clear that
\begin{equation}\label{}
    G_0'' = S'' G_0 S''^\dagger\ ,
\end{equation}
where $S''$ represents permutation $(1,2,3) \rightarrow (2,1,3)\,$,
that is
\begin{equation}\label{}
    S'' = \left( \begin{array}{ccc} \cdot & 1 & \cdot \\ 1 & \cdot & \cdot \\
    \cdot & \cdot & 1 \end{array} \right) \ .
\end{equation}
Hence a class of $G_0'' \ot \overline{G}_0''$--invariant states is
defined by
\begin{equation}\label{}
\rho'' = S'' \ot S''\, \rho\, S''^\dagger \ot S''^\dagger\ ,
\end{equation}
where $\rho$ is  $G_0 \ot \overline{G}_0$--invariant. The
corresponding matrix representation  of $\rho''$ has the following
form
\begin{equation}\label{2=3}
 \rho'' = \left( \begin{array}{ccc|ccc|ccc}
\rho_{11} & \cdot & \cdot &  \cdot & \rho_{15} & \rho_{16} &  \cdot & \rho_{18} &  \rho_{19} \\
\cdot & \rho_{22} & \rho_{23} & \cdot & \cdot & \cdot & \cdot &
\cdot & \cdot \\
\cdot & \rho_{32} &   \rho_{33} & \cdot & \cdot & \cdot & \cdot & \cdot & \cdot \\
\hline \cdot & \cdot & \cdot & \rho_{44} & \cdot & \cdot & \rho_{47}
& \cdot & \cdot \\
\rho_{51} & \cdot & \cdot & \cdot & \rho_{55} & \rho_{56} & \cdot &
\rho_{58} & \rho_{59} \\
\rho_{61} & \cdot & \cdot & \cdot & \rho_{65} &  \rho_{66}  & \cdot & \rho_{68} & \rho_{69} \\
\hline \cdot & \cdot
& \cdot &  \rho_{74} & \cdot & \cdot & \rho_{77} &    \cdot &  \cdot \\
\rho_{81} & \cdot
& \cdot &  \cdot & \rho_{85} & \rho_{86} &  \cdot &  \rho_{88} &  \rho_{89} \\
\rho_{91} & \cdot &  \cdot & \cdot &  \rho_{95} & \rho_{96} & \cdot
& \rho_{98} & \rho_{99}
\end{array} \right)\  .
\end{equation}
In particular one obtains the following representation of the
Horodecki state invariant under $G''_0$
\begin{equation}\label{}
{\rho_a}'' = S'' \ot S''\, \rho_a\, S''^\dagger \ot S''^\dagger\ ,
\end{equation}
that is,
\begin{equation}\label{RHO-23}
    {\rho_a}''\, =\, N_a\left(\begin{array}{ccc|ccc|ccc}
   a&\cdot&\cdot&\cdot&a&\cdot&\cdot&\cdot&a\\
   \cdot&a&\cdot&\cdot&\cdot&\cdot&\cdot&\cdot&\cdot\\
   \cdot&\cdot&a&\cdot&\cdot&\cdot&\cdot&\cdot&\cdot\\
   \hline
   \cdot&\cdot&\cdot&a&\cdot&\cdot&\cdot&\cdot&\cdot\\
   a&\cdot&\cdot&\cdot&b&c&\cdot&\cdot&a\\
   \cdot&\cdot&\cdot&\cdot&c&b&\cdot&\cdot&\cdot\\
   \hline
   \cdot&\cdot&\cdot&\cdot&\cdot&\cdot&a&\cdot&\cdot\\
   \cdot&\cdot&\cdot&\cdot&\cdot&\cdot&\cdot&a&\cdot\\
   a&\cdot&\cdot&\cdot&a&\cdot&\cdot&\cdot&a
   \end{array}\right) \ .
\end{equation}
Again, the characteristic feature of (\ref{2=3}) is that $\rho''$
has a direct sum structure $\rho'' = \rho_1'' \oplus \rho_2'' \oplus
\rho_3''$\, where the corresponding operators $\rho_k$ are supported
on $\mathcal{H}_k''$
\begin{eqnarray}
  \mathcal{H}_1'' &=& (S''\ot S'')\mathcal{H}_1 =  \rm{span}_{\,\mathbb{C}} \{\, |11\>,\ |23\>,\ |22\>,\ |32\>,\ |33\>\, \}\ , \nonumber\\
  \mathcal{H}_2'' &=&  (S''\ot S'')\mathcal{H}_2 = \rm{span}_{\,\mathbb{C}} \{\, |21\>,\ |31\>\, \}  \ , \\
  \mathcal{H}_3'' &=& (S''\ot S'')\mathcal{H}_3 =  \rm{span}_{\,\mathbb{C}} \{\, |12\>,\ |13\>\, \} \nonumber
  \ .
\end{eqnarray}
One easily finds for the partial transposition
\begin{equation}\label{2=3-Gamma}
 {\rho''}^\Gamma = \left( \begin{array}{ccc|ccc|ccc}
\rho_{11} & \cdot & \cdot &  \cdot & \cdot & \cdot &  \cdot & \cdot &  \cdot \\
\cdot & \rho_{22} & \rho_{32} & \rho_{15} & \cdot & \cdot &
\rho_{18} &
\cdot & \cdot \\
\cdot & \rho_{23} &   \rho_{33} & \rho_{16} & \cdot & \cdot & \rho_{19} & \cdot & \cdot \\
\hline \cdot & \rho_{51} & \rho_{61} & \rho_{44} & \cdot & \cdot &
\rho_{47}
& \cdot & \cdot \\
\cdot & \cdot & \cdot & \cdot & \rho_{55} & \rho_{65} & \cdot & \rho_{58} & \rho_{68} \\
\cdot & \cdot & \cdot & \cdot & \rho_{56} &  \rho_{66}  & \cdot & \rho_{59} & \rho_{69} \\
\hline \cdot & \rho_{81}
& \rho_{91} &  \rho_{74} & \cdot & \cdot & \rho_{77} &    \cdot &  \cdot \\
\cdot & \cdot
& \cdot &  \cdot & \rho_{85} & \rho_{95} &  \cdot &  \rho_{88} &  \rho_{98} \\
\cdot & \cdot &  \cdot & \cdot &  \rho_{86} & \rho_{96} & \cdot &
\rho_{89} & \rho_{99}
\end{array} \right)\  ,
\end{equation}
which is supported  the direct product of three subspaces
\begin{eqnarray}
  \widetilde{\mathcal{H}}_1''  &=& (S''\ot S'')\widetilde{\mathcal{H}}_1 = \rm{span}_{\,\mathbb{C}} \{\, |21\>,\ |23\>,\ |32\>,\ |33\>\, \} \ , \nonumber\\
  \widetilde{\mathcal{H}}_2'' &=&  (S''\ot S'')\widetilde{\mathcal{H}}_2 = \rm{span}_{\,\mathbb{C}} \{\, |12\>,\ |21\>,\ |13\>,\ |31\>\, \} \ , \\
  \widetilde{\mathcal{H}}_3'' &=& (S''\ot S'')\widetilde{\mathcal{H}}_3 =  \rm{span}_{\,\mathbb{C}} \{\, |11\>\, \} \nonumber
  \ .
\end{eqnarray}
It is evident that the analog of (\ref{HHHH}) is satisfied for
$\mathcal{H}_k''$ and $ \widetilde{\mathcal{H}}_k''$.

\section{Conlcusions}

We shown that the celebrated Horodecki state \cite{Pawel} belongs to
a class of states invariant under a commutative subgroup $G_0$ of
$U(3)$. Taking conjugate subgroups $G_0'$ and $G_0''$ we provided
another classes of invariant states. In particular we found
equivalent representations of the Horodecki state invariant under
$G_0'$ and $G_0''$, respectively (cf. formulae (\ref{RHO-12}) and
(\ref{RHO-23})). Interestingly, known entanglement witnesses
detecting PPT entangled state (\ref{RHO}) display $G_0$-invariance
(see \cite{W1,W2}). It should be clear that our discussion can be
immediately generalized from $3 \ot 3$ to  $d \ot d$ ($d$ arbitrary
but finite). Now, the maximal commutative subgroup of $U(d)$ defined
by (\ref{U-x}) gives rise to a number of subgroups corresponding to
$x_{k_1} = \ldots = x_{k_l}$. In particular using a subgroup defined
by $x_1=x_d$ one may introduce the  generalized Horodecki state in
$d \ot d$. We believe that our discussion opens new perspectives to
study symmetric states of composite quantum systems. It would be
interesting to generalize our analysis to multipartite case
\cite{multi1,multi2}.

\section*{Acknowledgments}

 This work was partially supported by
the Polish Ministry of Science and Higher Education Grant No
3004/B/H03/2007/33.


\begin{thebibliography}{1} \bibliographystyle{plain}

\bibitem{Pawel} P. Horodecki,  Phys. Lett. A {\bf 223}, 1
(1996).

\bibitem{HHHH} R. Horodecki, P. Horodecki, M. Horodecki and K.
Horodecki, Rev. Mod. Phys. {\bf 81}, 865 (2009).


\bibitem{Guhne} O. G\"uhne and G. T\'oth, Phys. Rep.  {\bf 474}, 1
(2009).

\bibitem{ISO} M. Horodecki and P. Horodecki, Phys. Rev. A {\bf 59},
4206 (1999).

\bibitem{Werner1} R.F. Werner, Phys. Rev. A {\bf 40}, 4277 (1989).

\bibitem{PPT-nasza} D. Chru\'sci\'nski and A Kossakowski, Phys.
Rev. A {\bf 74}, 022308 (2006).


\bibitem{Bruss} D. Bruss and A. Peres, Phys. Rev. A {\bf 61},
030301(R) (2000).

\bibitem{Djokovic}   D.Z. Djokovic,  {\em The checkerboard family of entangled states of two
    qutrits}, arXiv:0911.2797.

%\bibitem{QIT} M. A. Nielsen and I. L. Chuang, {\it Quantum computation
%and quantum information}, Cambridge University Press, Cambridge,
%2000.





\bibitem{CIRCULANT} D. Chru\'sci\'nski and A. Kossakowski,  Phys. Rev. A
{\bf 76}, 032308 (2007).

\bibitem{Brazylia}  F.E.S. Steinhoff and M.C. de Oliveira,
    {\em Families of bipartite states classifiable by the positive partial transposition
    criterion}, arXiv:0906.1297.


%\bibitem{Art} D. Chru\'sci\'nski  and A. Pittenger, J. Phys. A: Math. Theor. {\bf 41},
%385301 (2008).



%\bibitem{How} D. Chru\'sci\'nski and A. Kossakowski, J. Phys. A: Math. Theor. {\bf 41},
%145301 (2008).

\bibitem{W1} S. Yu and N. Liu, Phys. Rev. Lett, {\bf 95}, 150504
(2005).



\bibitem{W2} N. Ganguly and S. Adhikari, Phys. Rev. A {\bf 80}, 032331
(2009).


\bibitem{multi1} K.G.H. Vollbrecht and R.F.Werner, Phys. Rev. A {\bf 64},
062307 (2001).

\bibitem{multi2} D. Chru\'sci\'nski, A. Kossakowski,  Phys. Rev. A {\bf 73}, 062313
(2006); Phys. Rev. A {\bf 73}, 062314 (2006).




\end{thebibliography}
\end{document}